\documentclass[twocolumn,prb,reprint,superscriptaddress,floatfix,amsmath,amssymb,aps]{revtex4-2}

\usepackage[utf8]{inputenc}
\usepackage[caption=false,subrefformat=parens,labelformat=parens]{subfig}
\usepackage{graphicx}
\usepackage{float}
\usepackage{amsmath}
\usepackage{dcolumn}
\usepackage{bm}
\usepackage{xcolor}
\usepackage{xfrac}
\usepackage[colorlinks=true, allcolors=blue]{hyperref}
\usepackage{chemformula}
\usepackage{siunitx}

\def\be{\begin{equation}}       \def\ee{\end{equation}}
\def\bea{\begin{eqnarray}}      \def\eea{\end{eqnarray}}
\def\ba{\begin{array} }
\def\ea{\end{array} }
\def\bnum{\begin{enumerate} }
\def\enum{\end{enumerate}}

\def\=>{\Rightarrow}
\def\>{\rightarrow}

\def\eye2{Fathbb{I}}

\def\d0{\Delta_{0}}

\begin{document}
\title{\bf
The nematic susceptibility of the ferroquadrupolar metal TmAg$_2$ measured via the elastocaloric effect
}
\author{Elliott W. Rosenberg}
\affiliation{1) Geballe Laboratory for Advanced Materials and Department of Applied Physics, Stanford University, Stanford, CA 94305, USA}
\author{Matthias Ikeda}
\affiliation{1) Geballe Laboratory for Advanced Materials and Department of Applied Physics, Stanford University, Stanford, CA 94305, USA}
\author{Ian R. Fisher}
\affiliation{1) Geballe Laboratory for Advanced Materials and Department of Applied Physics, Stanford University, Stanford, CA 94305, USA}
\date{\today }
\begin{abstract}
    Elastocaloric measurements of the ferroquadrupolar/nematic rare-earth intermetallic TmAg$_2$ are presented. TmAg$_2$ undergoes a cooperative Jahn-Teller-like ferroquadrupolar phase transition at 5K, in which the Tm$^{3+}$ ion's local $4f$ electronic ground state doublet spontaneously splits and develops an electric quadrupole moment which breaks the rotational symmetry of the tetragonal lattice. 
    The elastocaloric effect, which is the temperature change in the sample induced by adiabatic strains the sample experiences, is sensitive to quadrupolar fluctuations in the paranematic phase which couple to the induced strain.  We show that elastocaloric measurements of this material reveal a Curie-Weiss like nematic susceptibility with a Weiss temperature of $\approx 2.7K$, in agreement with previous elastic constant measurements. Furthermore, we establish that a magnetic field along the c-axis acts as an effective transverse field for the quadrupole moments.    
\end{abstract}

\maketitle

\section{Introduction}
Intermetallic rare-earth compounds, in which the partially filled $4f$ orbitals associated with the rare-earth ions form a lattice of local electronic degrees of freedom, can provide models for studying cooperative phase transitions arising from microscopically well-understood multipole interactions ~\cite{quadhandbook}.  If the $4f$ electronic orbitals form a degenerate ground state, then the system (at low temperatures relative to any excited states) can be approximated as an sublattice of electronic multipolar moments coupled to non-$4f$ itinerant metallic degrees of freedom \cite{TransIs}. Materials such as PrPb$_3$, DySb, and YbRu$_2$Ge$_2$ all exhibit low temperature phase transitions in which the $4f$ sublattice plays a crucial role in driving multipolar phase transitions~\cite{morin1982magnetic,giraud1985multipolar,jeevan2006quasiquartet,rosenberg2019divergence,thalmeier2008multipolar}. TmAg$_2$ is also such a metal, in which the $4f$ ground state $\Gamma_5$ doublet spontaneously splits through undergoing $B_{1g}$ (i.e. $x^2-y^2$ symmetry) ferroquadrupolar order at 5K, breaking the $C_4$ rotational symmetry of the tetragonal crystal structure~\cite{TmAg2QO,morin1995specific}. This phase transition does not break translational symmetry, and so from a symmetry perspective is equivalent to the electronic nematic phase transitions that have been recently investigated in a variety of material families including the iron pnictide and chalcogenide superconductors~\cite{kasahara2012electronic,pnictide2,pnictide3,bohmer2017nematicity}. 

The effective Hamiltonian describing the low-energy properties of $4f$ ions incorporated in crystalline solids is understood in great detail. For cases in which the Kondo energy scale is vanishingly small~\cite{quadhandbook}, the $4f$ electronic wavefunctions are spatially localized, and so form a lattice of local electronic sites.    Each site has strong spin-orbit coupling which yields an electronic multiplet that is described by a total angular momentum number J. Due to the local nature of the multiplet, the crystal electric field (CEF) from the surrounding ligands acts as a weak perturbation, splitting the degeneracy of the spherical harmonic basis states.  Coupling between $4f$ orbitals of different lattice sites arises both from the interaction of the quadrupolar degrees of freedom with the lattice, and the generalization of the Ruderman–Kittel–Kasuya–Yosida (RKKY) interaction, mediated by the conduction electrons in metals~\cite{quadhandbook}. This provides an effective exchange interaction between the $4f$ sites, analogous to the exchange interaction that appears in magnetic Ising models.

Significantly, the low-energy states of these $4f$ electronic multiplets can exhibit degeneracies, the multiplicity of which is determined by the point-group symmetry and charge distribution surrounding the $4f$ ion.  For tetragonal materials with even angular momentum number J, $\Gamma_5$ doublets, (i.e. $E_g$ symmetry, which transforms like ($xz$, $yz$)), can be the ground state of the $4f$ system, as is the case for TmAg$_2$. At low enough temperatures (relative to the higher energy states determined by the crystalline effective field), thermal occupation of the excited $4f$ states can be neglected, and so materials with this ground state configuration are ideal candidates to compare to a pseudo-spin 1/2 Ising model in which sites are coupled via both the lattice and the metallic degrees of freedom~\cite{TransIs}. 

As shown in previous work (Ref.~\cite{TransIs}), in this restricted Hilbert space for TmAg$_2$, $B_{2g}$ strain (i.e. $\varepsilon_{xy}$) and magnetic field $H_z$ directed along the c-axis map precisely to transverse fields, and so this system maps to a transverse field Ising model with an additional coupling between the sites and the lattice~\cite{TmVO4Pierre}. Notably, a magnetic field has been shown to suppress this phase transition to 2K for an applied field of 3T, allowing exploration of a possible quantum phase transition if the phase diagram could be followed to lower temperatures~\cite{TmAg2QO}. However, as mentioned previously, it should be emphasized that TmAg$_2$ contains metallic degrees of freedom that interact with the local $4f$ Ising lattice, unlike the paradigmatic case of $4f$ ferroquadrupolar order in the insulating compound TmVO$_4$~\cite{TmVO4,TmVO4Pierre,TmVO4NMR}.  We might still expect to tune the ferroquadrupolar transition towards a metallic nematic quantum critical point via the application of a transverse field, but this assumes the metallic degrees of freedom don't couple strongly enough to the $4f$ sublattice to fundamentally change the low-temperature behavior of the total coupled system. The interactions between the metallic part of this material and the $4f$ sublattice are then of particular interest in the proximity of the potential quantum critical point.

TmAg$_2$ undergoes only one phase transition, a ferroquadrupolar phase transition occuring at 5K~\cite{TmAg2QO}. The local $4f$ electronic multiplets, which for the Tm$^{3+}$ ion harbor quadrupolar degrees of freedom, develop spontaneous electronic quadrupolar moments which collectively align at this phase transition. The macroscopic quadrupolar order parameter breaks the four-fold rotational symmetry of the lattice, and so the lattice distorts (i.e develops a spontaneous strain) in a similar fashion due to its coupling to the order parameter. Herein lies a crucial point: because strain, which represents the internal degree of freedom of the lattice to distort, is a rank two tensor quantity (strain is defined as the symmetric spatial derivative of a displacement vector), it couples bilinearly to the quadrupoles, also rank two, of the same symmetry. This coupling guarantees that for a finite applied strain (induced by external stresses) of the appropriate symmetry there will be a non-zero induced quadrupolar moment, and vice versa. Thus techniques in which the material experiences strain are directly applying the conjugate field for ferroquadrupolar phase transitions, and so are of considerable interest for materials like TmAg$_2$. Specifically, we use the AC elastocaloric effect technique, in which AC strains are induced in a sample, and the resulting temperature changes (in quasi-adiabatic conditions) arising from electronic entropy shifts are measured,  which recently was developed and applied in studying pnictide superconductors, and apply it to TmAg$_2$ in this paper~\cite{ikeda2019ac,ikeda2021elastocaloric}. 

The elastocaloric technique, described in detail in~\cite{ikeda2019ac}, probes strain derivatives of the entropy, analogous to the magnetocaloric effect which measures derivatives of the entropy with respect to magnetic field. Thus even in the disordered state of a material, at temperatures above the relevant phase transition temperature,  if a finite order parameter is induced by a strain of the relevant symmetry, then there will be a non-zero change in the entropy and hence a potentially measurable change in the temperature of the sample. In fact for phase transitions in which the terms (to second order) in the Ginzburg-Landau free energy can be written as:

\begin{equation}
F=a(T)\psi_{\Gamma}^2+\lambda\psi_{\Gamma}\varepsilon_{\Gamma} +\frac{1}{2}C_{\Gamma}\varepsilon_{\Gamma}^2+...
\end{equation}

where $\psi_{\Gamma}$ is the order parameter, $\varepsilon_{\Gamma}$ is the corresponding strain of the same irreducible representation (irrep) as the order parameter ($\Gamma$), $\lambda$ is the strain-order parameter coupling constant, and $C_{\Gamma}$ is the appropriate elastic constant for the irrep of strain, the elastocaloric effect, which is the derivative of the temperature change of the sample with respect to an induced strain, can be shown to probe the order parameter-strain susceptibility of the material:

\begin{equation}
\frac{\partial T}{\partial \varepsilon_{\Gamma}}=-\frac{T}{C_p}\frac{\partial S}{\partial \varepsilon_{\Gamma}}=\frac{T\lambda^2\varepsilon^o_{\Gamma}}{C_p}\frac{d\chi_{\Gamma}}{dT}
\end{equation}

where C$_p$ is the heat capacity of the sample, $\chi_{\Gamma} = \frac{d\psi_{\Gamma}}{d\varepsilon_{\Gamma}}$, and $\varepsilon^o_{\Gamma}$ is the offset strain of the sample, discussed shortly. Note that this equation features partial derivatives with respect to strains, and so each strain tensor element induced will have an associated elastocaloric effect. Since strain tensor elements cannot be induced in isolation experimentally, (a uniaxial stress will induce mixtures of the relevant different irreps of strain based on the corresponding Poisson ratios)~\cite{symmandasymm}, an elastocaloric measurement will probe an admixture of strain derivatives of the entropy. However Eq.~(1)  is only relevant for anti-symmetric strains (symmetric strains will have dominant linear terms in the free energy without bilinear couplings), and so measurements that display a linear relationship with the applied offset strain $\varepsilon^o_{\Gamma}$ (thus changing sign through zero strain) can be reasonably assumed to arise from nematic susceptibilies.


  For materials like the Ba(Fe$_{1-x}$Co$_x$)$_2$As$_2$ family, in which $\varepsilon_{B_{2g}}$ strain plays a dominant role in the entropy landscape,  the elastocaloric effect has been shown to be a sensitive probe in determining the thermodynamic nematic susceptibility above the structural phase transition \cite{ikeda2021elastocaloric}. We show for the case of the ferroquadrupolar TmAg$_2$, in which the B$_{1g}$ quadrupole orders at 5K, that the elastocaloric technique can provide an isolated measurement of the B$_{1g}$ nematic susceptibility, as our measurement aligns well with previous elastic constant measurements. We also demonstrate by applying magnetic fields with both heat capacity and elastocaloric measurements, that the ferroquadrupolar phase and nematic susceptibility in TmAg$_2$ are well described by a semi-classical mean-field transverse-field Ising phase diagram in which a magnetic field oriented along the c-axis $H_z$ acts an effective transverse field. 

\subsection{Model Hamiltonian for $4f$ degrees of freedom}
We first introduce a model $4f$ Hamiltonian for TmAg$_2$ neglecting metallic degrees of freedom (besides the inclusion of a interaction between $4f$ sites mediated via conduction electrons treated at a mean-field level), for which the CEF spectrum has a ground state non-Kramer's doublet which maps precisely to the spin 1/2 Ising model.

TmAg$_2$ is a rare-earth intermetallic material with tetragonal crystal structure at room temperature (space group I$_4$/mmm), and belongs to the MoSi$_2$ structure type~\cite{TmAg2QO,morin1995specific}. The $4f$ electrons of the trivalent Tm$^{+3}$ ion form local electronic multiplets with a total angular momentum quantum number J=6. The 2J+1=13 states are split by the surrounding crystalline electric field (CEF), and heat capacity and inelastic neutron scattering measurements provide evidence that the ground state is a $\Gamma_5$ doublet composed of $a|\pm 5\rangle + b|\mp 3\rangle +c|\pm 1\rangle$ states(where $|i\rangle$ denotes $m_j=i$.)~\cite{TmAg2QO}. The nearest excited state is a $\Gamma_1$ singlet (composed of $|\pm 4\rangle$  and $| 0\rangle$ states) approximately 14K higher in energy~\cite{TmAg2QO}. Most notably, this is the only metallic $4f$ tetragonal material known to undergo solely a continuous ferroquadrupolar phase transition, (coupled with an tetragonal to orthorhombic structural phase transition.) It occurs at 5K with an $O_2^2$ ($B_{1g}$ symmetry) quadrupolar order parameter. This phase transition is driven by local electronic degrees of freedom to break the rotational symmetry of the lattice, and hence is an example of an electronic nematic phase transition. No other phase transitions have been observed in this material above 1.5K.

The operators $O^2_2$, $J_z$ (and the $B_{2g}$ Stevens operator $P_{xy}$) map onto the the standard Pauli matrices $\sigma_i$ when applied to the restricted subspace of the ground state $\Gamma_5$ doublet~\cite{TransIs}. They obey the same canonical commutation relations, and so at low enough temperatures relative to the CEF excited states, and ignoring any complications from the metallic surrounding lattice, $H_z$ should theoretically serve as an effective transverse field to the spontaneous nematic B$_{1g}$ order, and hence can tune the $4f$ sublattice to quantum criticality. 


A model Hamiltonian, considering applied magnetic fields along the c-axis, can then be written down to describe the physics of the $4f$ sites in this material, where only a single site is considered through a mean-field approximation:
\begin{equation}
\begin{split}
H_{4f}=&H_{CEF}-B_2^2 \varepsilon_{B_{1g}}O^2_2-K_{B_{1g}}\langle O^2_2 \rangle O^2_2 - g_zJ_zH_z \\
\label{ModelHam}
\end{split}
\end{equation}
The nonperturbative H$_{CEF}$ describes the crystal electric field from the surrounding atoms, and as previously stated produces a ground-state $\Gamma_5$ doublet with an excited singlet at $\Delta_1 = 14K$, with other excited states 50K and higher~\cite{TmAg2QO}. The second term is the bilinear magnetoelastic coupling of the B$_{1g}$ quadrupole operator $O^2_2$  ($J_x^2 -J_y^2$)  to the corresponding lattice strain (induced or spontaneous) $\varepsilon_{B_{1g}}=\frac{1}{2}(\varepsilon_{xx}-\varepsilon_{yy})$, and the following term is a mean field quadrupole-quadrupole interaction term originating from the RKKY interaction mediated by the conduction electrons.  Note that this Hamiltonian will produce the bilinear terms like those described in the free energy model in Eq.~(1), and hence the quadrupole-strain susceptibility $\frac{d\langle O^2_2 \rangle}{d\varepsilon_{B_{1g}}}$ should be measurable with the AC elastocaloric technique if it induces $B_{1g}$ strain.

To model the effects of strain on this material, the operators in this Hamiltonian were applied in the full 13 dimensional Hilbert space of the single site $4f$ electronic multiplet, and Eq. 3 was self-consistently diagonalized to obtain the thermal expectation value of the quadrupolar operator. Quantities like the quadrupolar order parameter and the heat capacity are derived from the resulting free energy and shown respectively in Figures 2 and 3 as functions of temperature.

\begin{figure}
\centering
\includegraphics[width=0.5\textwidth]{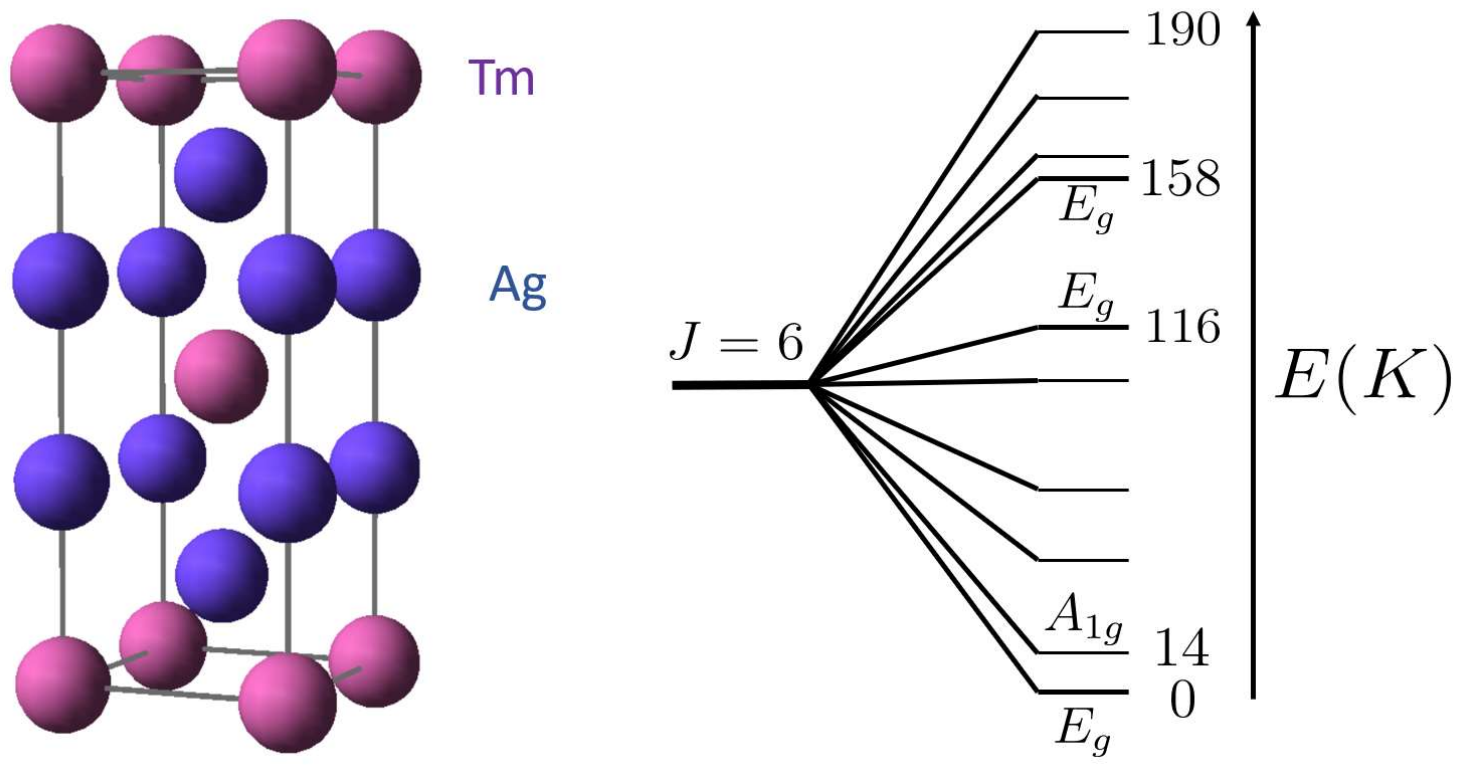}
\caption{{\bf Crystal structure and $4f$ CEF spectrum for the Tm$^{3+}$ ion.} The crystal structure is shown on the left, belonging to the space group I4/mmm with tetragonal MoSi$_2$ structure type. The silver atoms that surround Tm$^{3+}$ ions produce a crystalline electric field (CEF) with a D$_{4h}$ point group symmetry, which splits the 13 fold degeneracy of the $4f$ electronic multiplet (J=6), with previous neutron scattering measurements indicating an E$_g$ ground state and depicted excited states~\cite{TmAg2QO}.
}
\end{figure}

\label{CEF}

\begin{figure}
\centering
\includegraphics[width=0.8\textwidth]{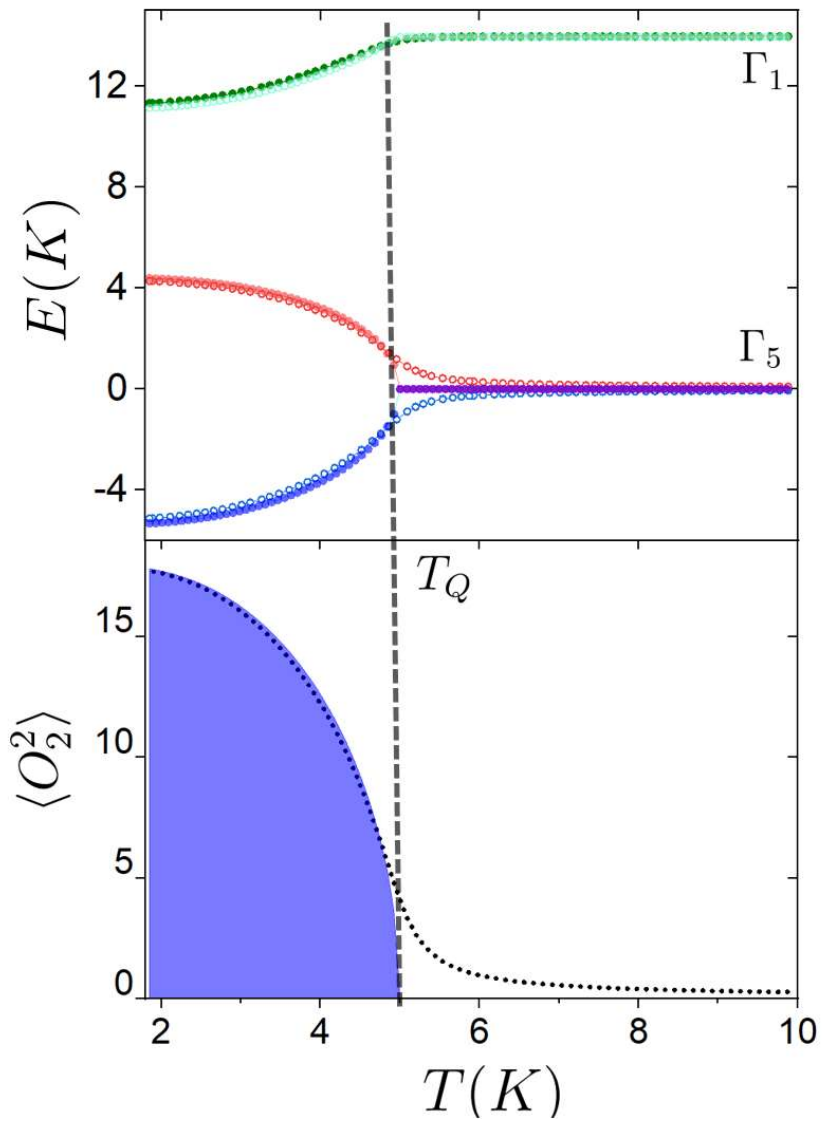}
\caption{{\bf Schematic of $4f$ ground state splitting and developing a quadrupolar order parameter.}  The top panel depicts a schematic of the three lowest energy levels of the $4f$ electronic multiplet, in which the ground state doublet splits spontaneously at 5K as each site develops a quadrupolar order parameter, the magnitude of which is shown in the bottom panel. The lighter color symbols in the energy level diagram correspond to the energies of a site experiencing strains in the B$_{1g}$ symmetry channel (similar in magnitude to the strains that were input to best match the experimental heat capacity shown in Figure 3), and the corresponding smearing of the transition manifests in a finite order parameter above 5K, as shown by the dotted line in the bottom panel. }
\label{OP}
\end{figure}
\begin{figure}
\centering
\includegraphics[width=0.4\textwidth]{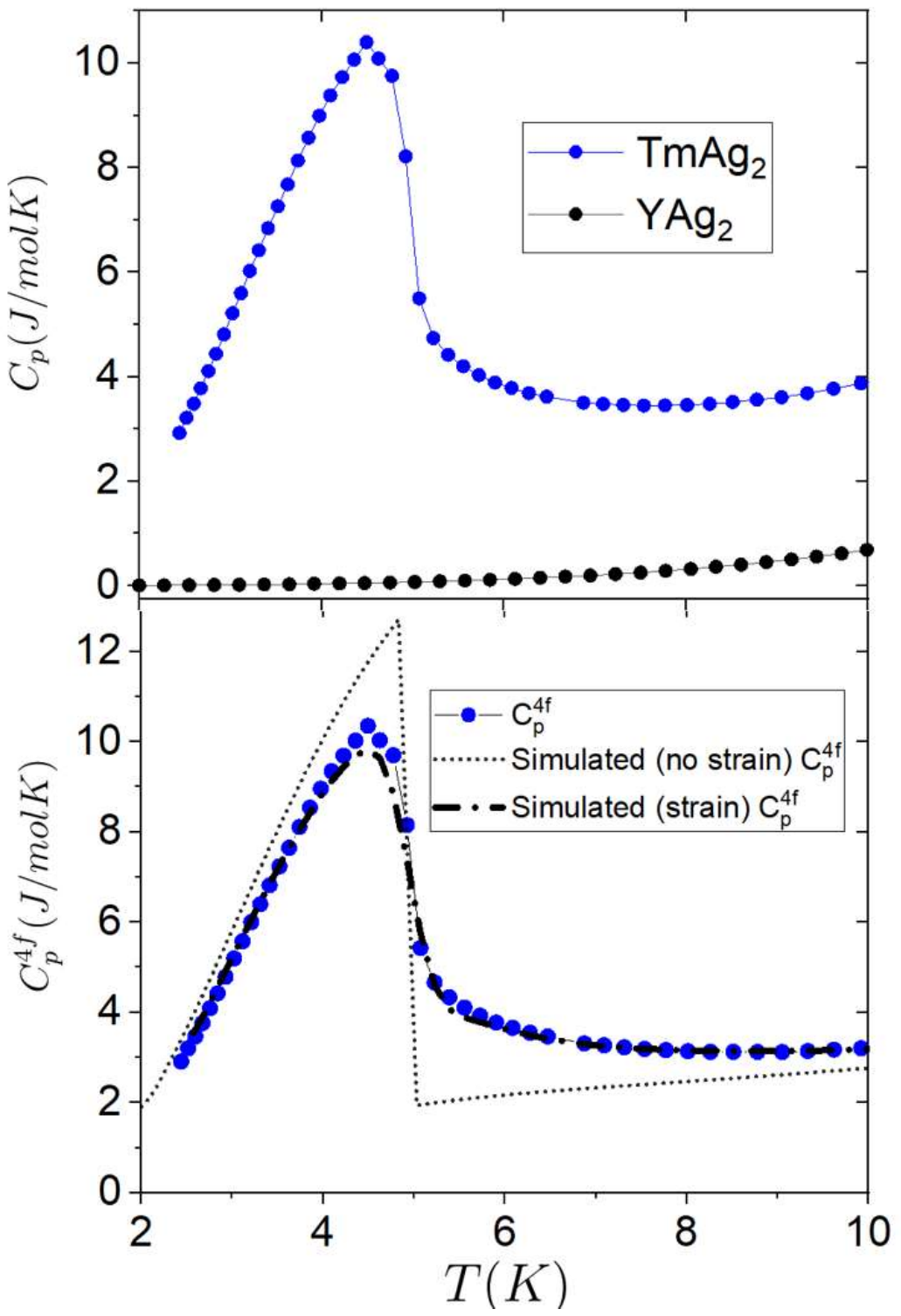}
\caption{{\bf \bf $4f$ Heat capacity of TmAg$_2$ compared to simulations with "built-in" strains.} The top panel depicts the measured heat capacity for both TmAg$_2$ and YAg$_2$. The bottom panel displays the difference of heat capacities of the two materials C$_p^{4f}$, denoted by blue dots. The small black dots are the results of simulated heat capacity just considering the CEF energy spectrum and a mean-field quadrupolar interaction term.  The black dashes are simulated heat capacity results which also include three $B_{1g}$ strains as fit parameters to account for the inherent broadening even in the nominally free-standing sample.  The phase transition temperatures in the simulations were determined by the mean-field interaction constants $B^2/C +K$, which were chosen so the system had a critical temperature $T_Q=5K$. This simulated heat capacity is the linear combination: $0.634C_p(\varepsilon_1)+0.23C_p(\varepsilon_2)+0.136C_p(\varepsilon_3)$ where $\varepsilon_1=0.007\%$, $\varepsilon_2=0.1\%$, and $\varepsilon_3=0.7\%$. These values were chosen to mimic the distribution of strains likely present in the material even in the absence of induced strains.}
\label{Cpandsim}
\end{figure}

\section{Results and Discussion}

\subsection{Heat Capacity in zero magnetic field}

\begin{figure}
\centering
\includegraphics[width=0.5\textwidth]{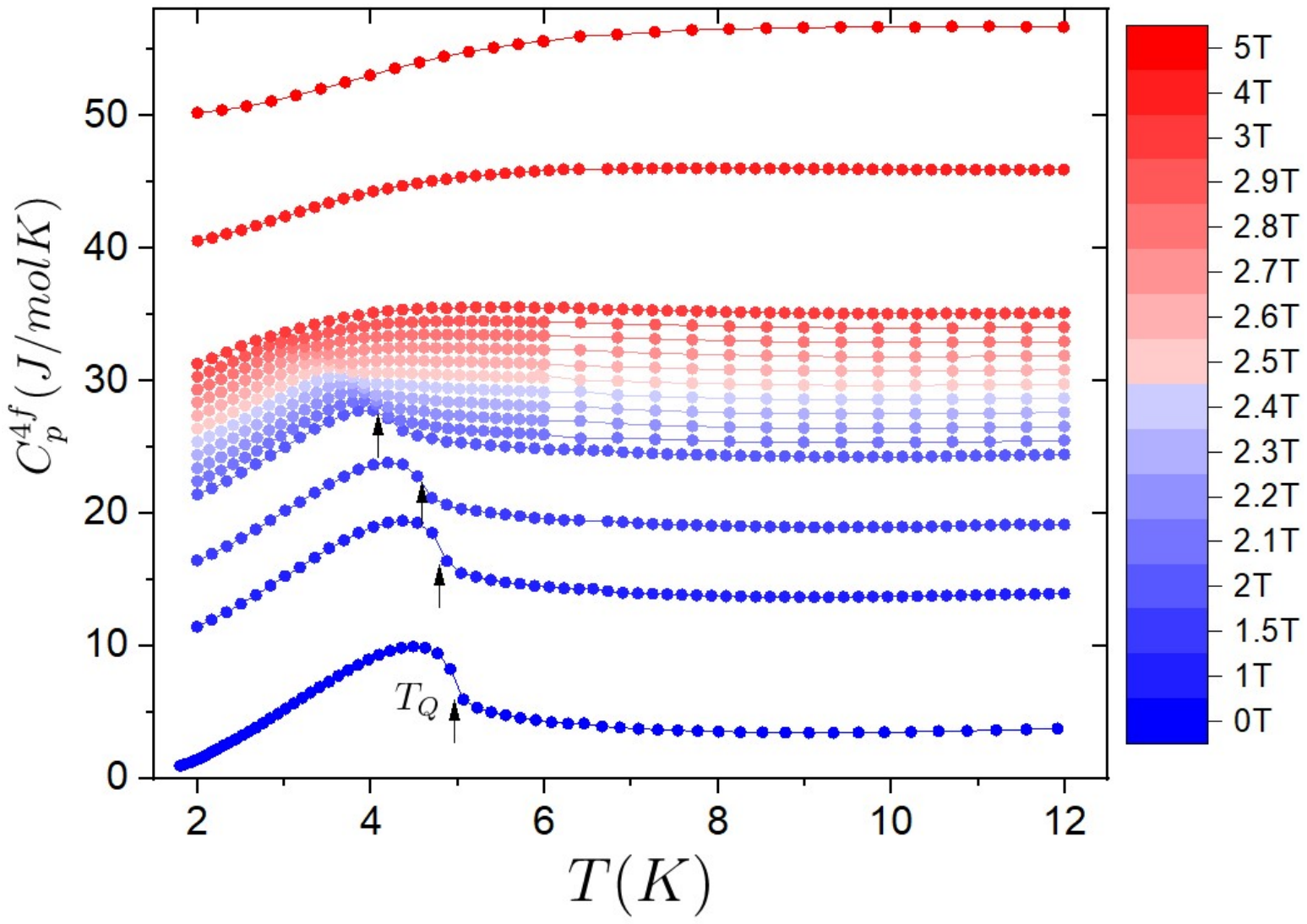}
\caption{{\bf Variation of the temperature dependence of the heat capacity with applied magnetic field.} Data are shown for fields ranging from 0-5T, with vertical offsets proportional to the applied magnetic field for clarity (no offsets for $H=0T$). T$_Q$ is marked by a black arrows, showing its progressive suppression with increased magnetic field H$_z$}
\label{Cpinfield}
\end{figure}

\begin{figure}
\centering
\includegraphics[width=0.5\textwidth]{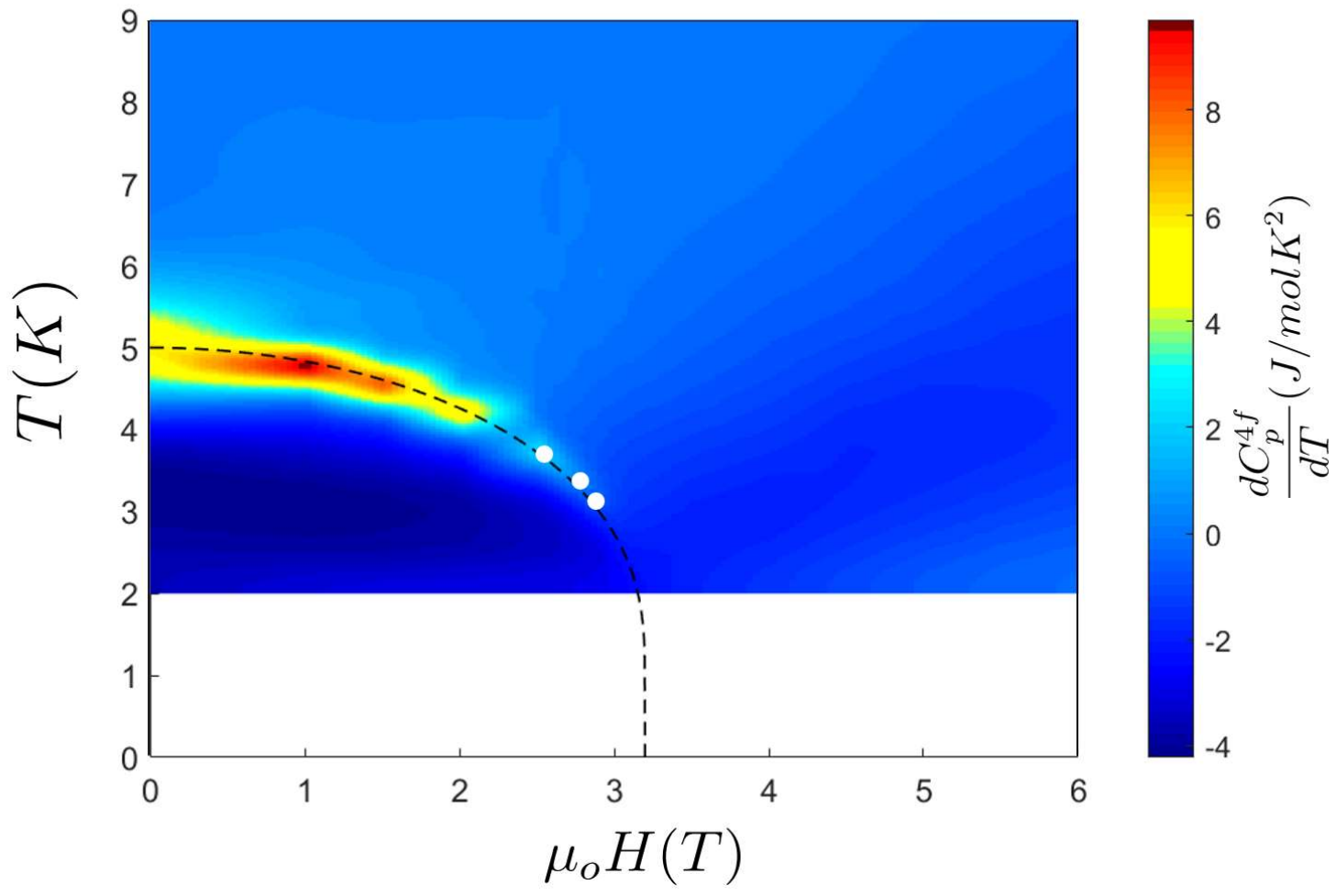}
\caption{{\bf Interpolated color map of the temperature derivative of the $4f$ heat capacity in the H-T plane.} The dotted line shows a semi-classical transverse field Ising fit of a phase boundary with $T_Q =5K$ at zero field and $H_c=3.2 \pm 0.1T$ at 0K. White dots depict the peaks of $dC_p/dT$ which are difficult to see given the color shading.}
\label{HCcolor}
\end{figure}
Heat capacity measurements were performed for several samples with different levels of disorder evident in the broadening of the phase transition. The least disordered is presented here in the top panel of Figure \ref{Cpandsim}. The ferroquadrupolar phase transition manifests in a mean-field step at 5K, with is consistent with a continuous phase transition mediated through long-range interactions (here the lattice) increasing the effective dimension above the upper critical dimension for the Ising model~\cite{TmVO4Pierre}. Note this is in contrast to the heat capacity of YAg$_2$ shown in the same panel, which only displays a similar phonon contribution as the Y ion has an empty $4f$ shell.  Subtracting the two measured heat capacities will isolate the $4f$ contribution which we call C$_p^{4f}$. This quanitity is displayed in the panel below, and it can be noted that the mean-field step signature is smeared, and that there is a sizable contribution to the heat capacity above 5K. To attempt to explain this we included results from the previously described simulation (thick dashed line) with ``built-in" strains that could represent macroscopic strains locked in the growth process as well as microscopic strains induced by impurities/ dislocation. Even from the calculation without these strains included (thin dotted line), the excited CEF states clearly provide a contribution to the heat capacity at temperatures above 5K.

\subsection{Heat capacity in magnetic fields}
The $C_p$ data from 0T to 5T is shown in Figure \ref{Cpinfield}, which shows a smooth suppression of $T_Q$ with magnetic field, as seen by the arrows. These traces (along with higher field traces) were interpolated in both field and temperature, and the numerical derivative was taken to create the color plot shown in Figure ~\ref{HCcolor}. The derivative marks the ferroquadrupolar phase transition quite well as a function of magnetic field down to around 3K, where the signature becomes too small in magnitude compared to the background to track accurately. The phase transition boundary is consistent with a semi-classical transverse field Ising phase boundary with $T_Q(0)=5K$ and $H_c=3.2\pm0.1T$. As the magnetic field was increased to progressively higher fields, a Schottky feature became evident. This feature is consistent with an energy gap that is linearly proportional to the applied magnetic field, which is expected for the energy spectrum of an isolated magnetic doublet.

\subsection{Elastocaloric measurements in zero magnetic field}

\begin{figure}
\centering
\includegraphics[width=0.5\textwidth]{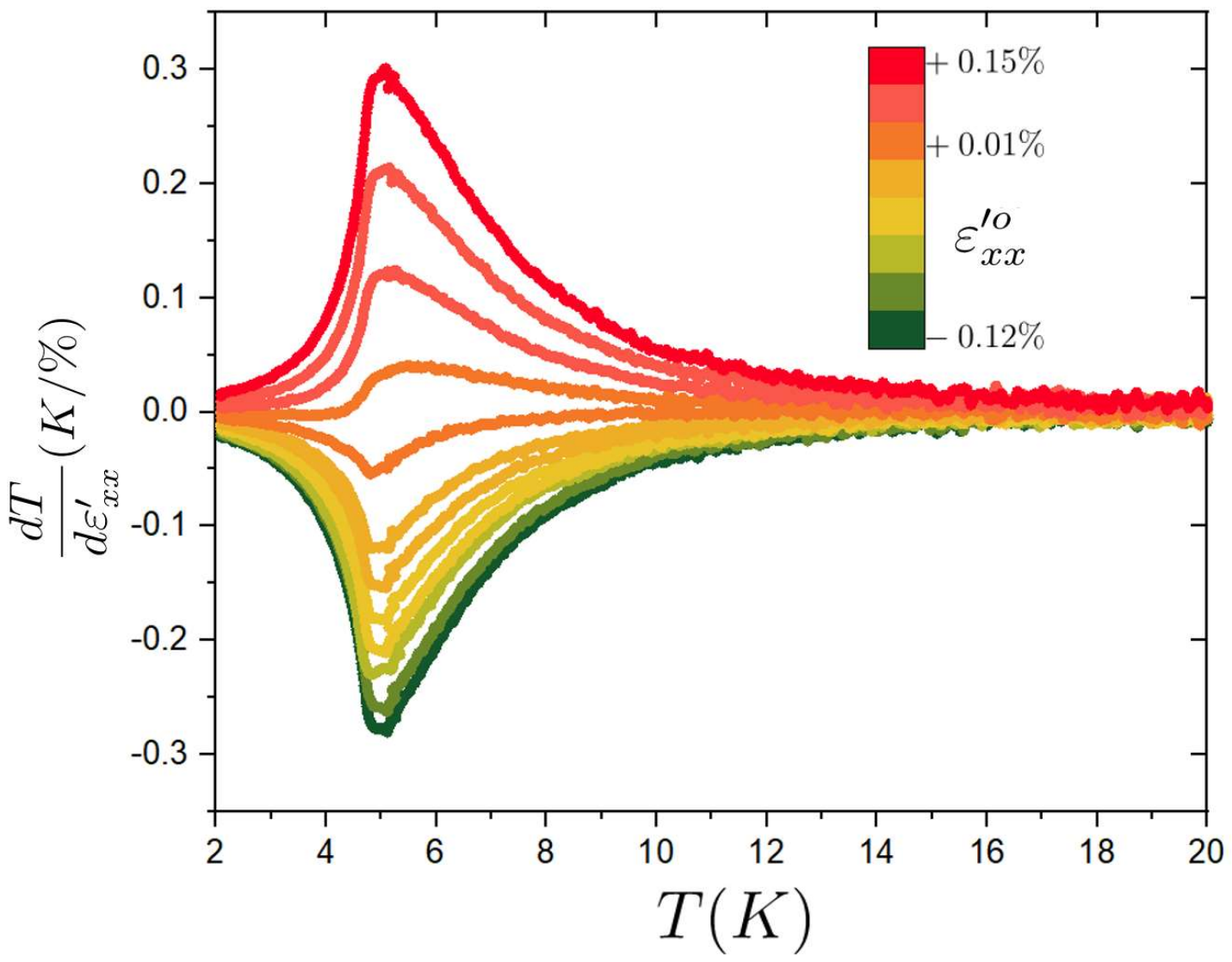}
\caption{{\bf Elastocaloric effect at different offset strains}  Plotted is the  magnitude of the temperature oscillations induced from a small AC strain ($\approx 0.01\%$) the sample experienced, which manifests from the elastocaloric effect. Uniaxial AC stress was applied along (1 0 0) while sweeping temperature from 20K to 2K,  for different DC stresses which induced strains $\varepsilon'_{xx}$ ranging from 0.15\% (in red) to -0.12\% (in green). 
}
\label{ECtraces}
\end{figure}

\begin{figure}
\centering
\includegraphics[width=0.5\textwidth]{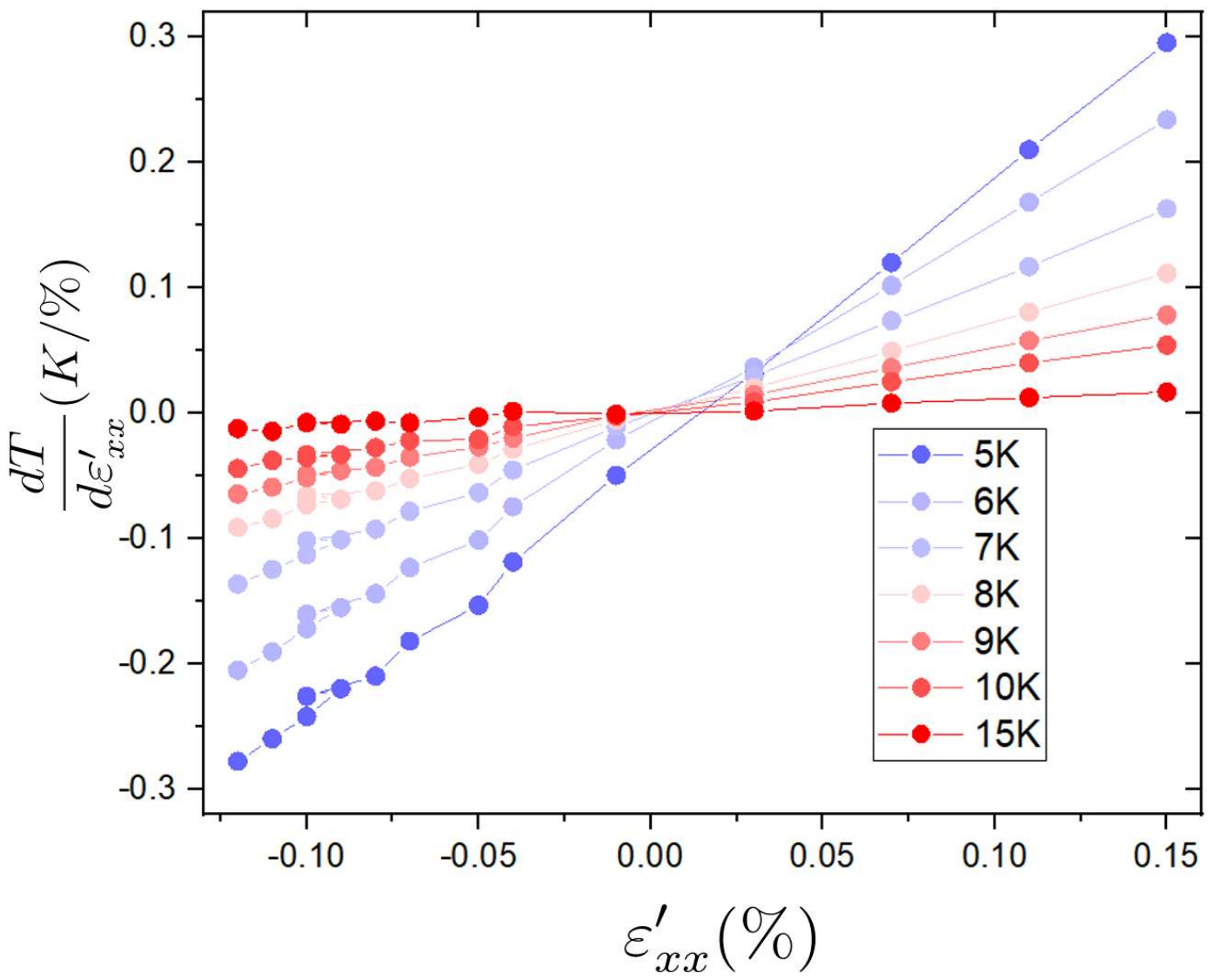}
\caption{{\bf Elastocaloric effect vs. DC strain}  The elastocaloric effect is plotted against the induced DC strain for different temperatures. The slope increases as the sample is cooled, indicating the nematic susceptibility is increasing towards 5K. The data is linearly proportional to the applied DC strain at all temperatures measured, indicating the DC and AC strains induced were well within the perturbative regime of the material.}
\label{ECvsDC}
\end{figure}

As shown via free energy arguments made previously, the elastocaloric effect  is a sensitive probe of fluctuations which couple to the induced strain that the sample is experiencing due to externally applied stresses. In particular, for a nematic/quadrupole-strain susceptibility of the material, the elastocaloric effect in the disordered phase will measure a signal $\big(\frac{dT}{d\varepsilon}\big)$ proportional to the strain multiplied by the temperature derivative of a nematic susceptibility (of the symmetry channel of the induced strain) as described in Eq. 2.


In the case of the elastocaloric measurements on TmAg$_2$ that appear in this work, the relevant nematic/quadrupole-strain susceptibility that is measured is the $B_{1g}$ susceptibility, $\chi_{B_{1g}}$. From the Hamiltonian in Eq. 3, a Curie-Weiss form can be expected for the $B_{1g}$ quadrupole-strain susceptibility:

\begin{equation} 
\chi_{B_{1g}}=\frac{d\langle O^2_2 \rangle}{d\varepsilon_{B_{1g}}}=\frac{(B^2_2)^2Q^2}{T-K_{B_{1g}}Q^2}
\end{equation}



where Q is the size of the saturated quadrupolar moment of the ground state. The elastocaloric effect $\frac{dT}{d\varepsilon'_{xx}}$ is displayed in Figure \ref{ECtraces}, with symbol colors ranging from red to green representing tensile to compressive induced DC strains. Within the range of the DC strains measured the elastocaloric signal remains linearly proportional to strain at all temperatures, signifiying the above equation accurately describes the data,and that effects arising from $A_{1g}$ symmetry strains have minimal effect on $\frac{dT}{d\varepsilon_{xx}}$. The response appears to diverge as temperature approaches the phase transition at 5K, as expected if the nematic susceptibility is Curie-Weiss like.

\begin{figure}
\centering
\includegraphics[width=0.5\textwidth]{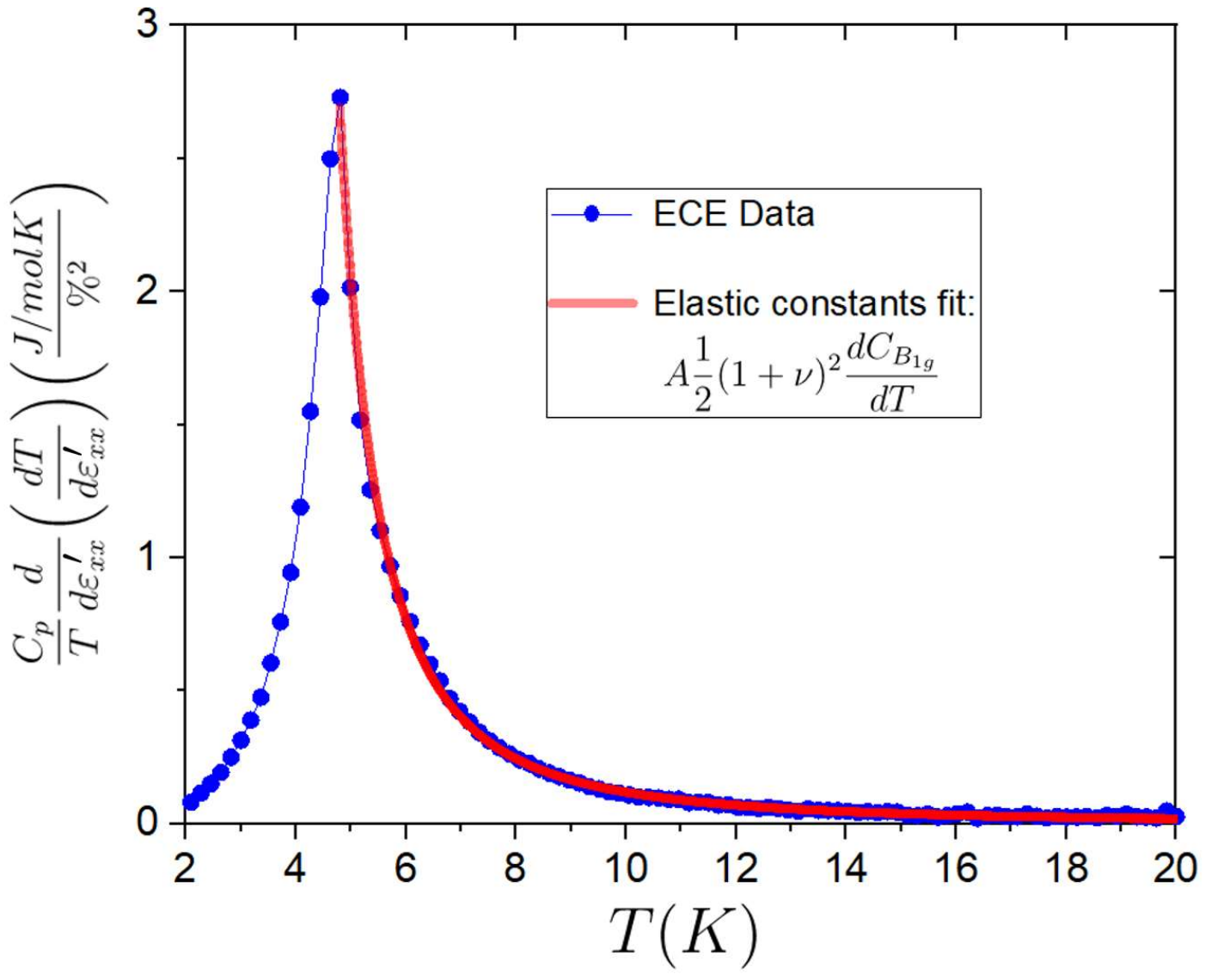}
\caption{{\bf Comparison of ECE with elastic constants} The blue points represent the temperature derivative of the nematic susceptibility obtained from the ECE. The red line is the temperature derivative of the B$_{1g}$ elastic constant, measured previously in~\cite{TmAg2QO},  multiplied by a proportionality constant dependent on the in-plane Poisson ratio which accounts for the softening of the lattice that manifests in a more sharply diverging ECE, and for the unknown strain transmission ratio ($\varepsilon'_{xx}$ vs. $\varepsilon_{xx}$) of this experimental setup, assumed to be temperature independent.}
\label{ECcompare}
\end{figure}

In Figure \ref{ECvsDC} the EC effect is plotted for different temperatures against the induced DC strains. The slope of this data was extracted at fixed temperatures, and to determine the nematic susceptibility that quantity was normalized by the ratio $C_p(T)/T$ to obtain $\frac{d\chi}{dT}$ as shown in Eq.~(2). This quantity is plotted in Figure \ref{ECcompare} against a quantity proportional to the temperature derivative of $B_{1g}$ elastic constants data measured by Ref.~\cite{TmAg2QO}. The proportionality constant is composed of two factors, one which is an arbitrary constant A that takes into account the unknown strain transmission rate of the elastocaloric measurements.  The second is given by $\frac{1}{2}(1+\gamma)^2$, which depends on the temperature dependent in-plane Poisson ratio $\gamma$ of the sample, is the ratio of the approximated $\varepsilon_{xx}^2$ to the actual $\varepsilon_{B_{1g}}^2$ induced. The temperature derivative of the nematic susceptibility as obtained from the elastocaloric effect vs. elastic constants softening nearly exactly align with a proposed Poisson ratio which softens from a room temperature value of 0.3 to the expected value of 1 at the phase transition. This provides further evidence that the elastocaloric measurement is dominated by the B$_{1g}$ quadrupolar susceptibility, as contamination from other symmetry modes would provide a background or temperature dependence which deviates from the elastic constant measurement. The constant A (proportional to the strain transmission ratio squared) provides a best fit when around 0.2, which indicates a strain transmission ratio of approximately 50\%. This is well within the expected strain transmission ratio of a hard relatively thick (60$\mu$m) sample at low temperatures.  Thus the nematic susceptibility as deduced from elastocaloric measurements of this material is consistent with that obtained from elastic constants. We note that both give an estimated Weiss temperature $T^*=K_{B_{1g}}Q^2=2.7K$ (taking into account the softening of the lattice in the elastocaloric effect, which artificially increases $T^*$), implying the quadrupolar interaction terms (that drive the ferroquadrupolar phase transition) which originate from the conduction electrons are nearly equal in magnitude to those that are mediated by the lattice.

\subsection{Field dependence of elastocaloric measurements}
Applying a magnetic field along the c-axis to the material in the disordered state will induce a gap $\Delta_H$ between the two states of the ground state doublet which will be proportional to the strength of the field. This will result in a reduced nematic susceptibility, and if the Hamiltonian in Eq.~(3) is considered only in this restricted Hilbert space, $\chi_{B_{1g}}$ is governed by the equation:
\begin{equation}
\chi_{B_{1g}}=\frac{d\langle O^2_2 \rangle}{d\varepsilon_{B_{1g}}}=\frac{(B^2_2)^2Q^2\tanh(\frac{\Delta_H}{k_bT})}{T-K_{B_{1g}}Q^2\tanh(\frac{\Delta_H}{k_bT})}
\end{equation}
This is the equation of the longitudinal order parameter susceptibility in the presence of an applied transverse field which induces a gap in the paranematic phase of magnitude $\Delta_H$.
\begin{figure}
\centering
\includegraphics[width=0.5\textwidth]{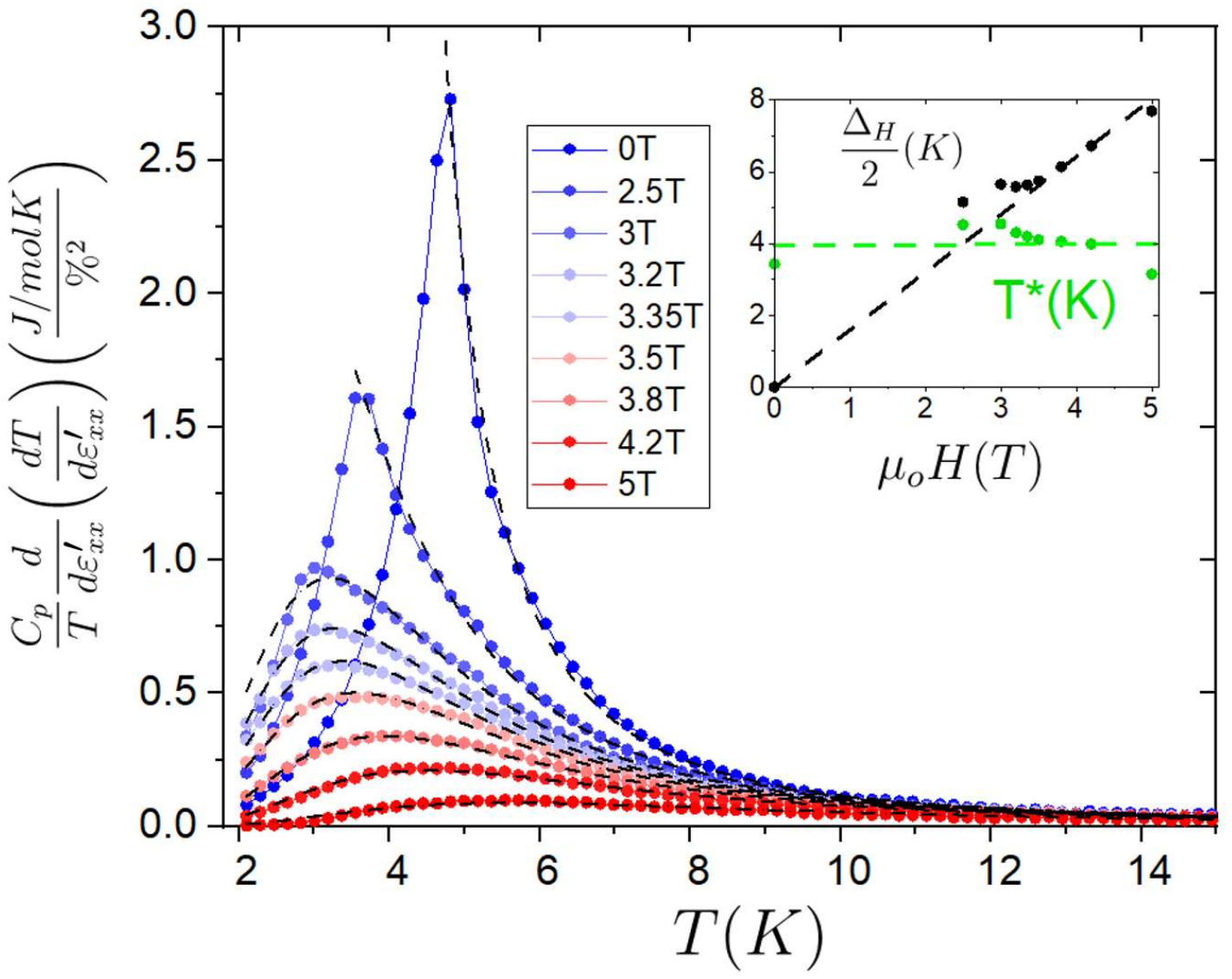}
\caption{{\bf Temperature dependence of the strain derivative of the elastocaloric effect for different magnetic fields.}  The strain derivative of the elastocaloric data (normalized by the fraction $\frac{T}{C_p}$ to obtain the temperature derivative of the nematic susceptibility) is plotted for magnetic fields applied along the c-axis from 0T (blue) to fields to near the critical field of $3.32T \pm 0.1T$ (light blue), to fields above the critical field (red). The black dashed lines represent best fits using the temperature derivative of Eq.(6) with the multiplying constant fixed at the zero-field value, and the varied parameters $\Delta_H(K)/2$ and $T^*(K)$ are plotted in the offset panel vs. magnetic field. }
\label{ECsus}
\end{figure}
The temperature derivative of the susceptibility as obtained from elastocaloric measurements is plotted in Figure \ref{ECsus}, with magnetic fields ranging from 0T to 5T, with higher density of magnetic fields measured near the critical field of $3.2\pm0.1$T to investigate possible consequences of quantum critical fluctuations. The results are described by Eq.~(5), with the fits at each field displayed as black dashed lines. These fits are quantitatively accurate at fields even above the critical field, although we note that in general the data is overdetermined by the free parameters in Eq.~(5), so we decided to keep $\lambda(B^2_2Q)^2$ fixed at the zero-field value, and let $\Delta_H$ and $T^*=K_{B_{1g}}Q^2$ vary as a function of field. As can be seen (inset to Figure 9), $T^*$ is approximately independent of field, while $\Delta_H$ displays nearly linear behavior with magnetic field, as expected from a simple polarized doublet ground state model. On closer inspection, deviations from these semi-classical expectations might be evident near the critical field (at which $T_Q$ is completely suppressed). However, the present sparse data density and sample quality preclude drawing any conclusions about whether these are signatures of quantum criticality. Nonetheless at a qualitative level, the elastocaloric effect in the paranematic phase of TmAg$_2$ is well explained via the quadrupole-strain susceptibility obtained from a transverse-field Ising Hamiltonian, with magnetic field along the c-axis acting as a transverse field to the B$_{1g}$ quadrupole order.

\section{Discussion}

We have established through elastocaloric and heat capacity measurements that the H-T phase diagram of TmAg$_2$ is consistent with a semi-classical transverse-field Ising model, with magnetic field along the c-axis acting as an effective transverse field to the $B_{1g}$ quadrupolar order parameter. 

This motivates the study of the thermodynamic susceptibilies and transport in the proximity of the completely suppressed ferroquadrupolar phase transition. While potential kinks/non-linearities in the fitting parameters of the thermodynamic nematic susceptibility were seen in the proximity of the critical field, the overdetermined nature of the fits makes this only a suggestive signature at best.  Rather, the more well-established experimental trend is that the quadrupole-strain susceptibility followed the expected form for a susceptibility in an spin 1/2 Ising model with magnetic field acting as an transverse field over the entire field and temperature range studied.
Further lower-temperature measurements are clearly motivated, performed using crystals with higher RRRs or sharper heat capacity features to resolve whether any consequences of quantum criticality are evident in transport or thermodynamic properties. 

We nevertheless want to emphasize the novelty of both the application of the elastocaloric effect on a $4f$ intermetallic material, as well as investigating  TmAg$_2$ in the context of nematic quantum criticality. The AC elastocaloric effect has recently been developed and applied to investigate an iron-based superconductor~\cite{ikeda2021elastocaloric}, as well as explore antiferroquadrupolar order in DyB$_2$C$_2$ ~\cite{ye2022elastocaloric}. We have shown with this work that it is also a sensitive probe of low-temperature ferroquadrupolar phase transitions, and uniquely suited to study materials with couplings to anti-symmetric strains. 
 For TmAg$_2$ we have shown that a c-axis magnetic field is consistent with functioning as a tuning parameter to obtain a quantum phase transition. As discussed in the introduction, this implies that B$_{2g}$ strain, theoretically an analogous transverse field along with magnetic field in the context of the $4f$ ground state, could also be a tuning parameter to obtain a quantum phase transition.  A strain-tuned quantum phase transition in TmAg$_2$ would be the first known case of a QCP tuned by \textit{anti-symmetric} strain, and motivates more strain-based measurements and developing different low-temperature experimental setups which allow larger strains.

\section{Methods}
TmAg$_2$ single crystals were grown from a melt inside tantalum crucibles. Stoichiometric amounts of thulium and silver were loaded into a crucible with one end open, and these crucibles were then capped and sealed in a tetra-arc furnace in an argon atmosphere, which was pumped and purged numerous times to reduce contamination. These Ta crucibles were sealed under vacuum in quartz tubes, then heated in a box furnace to 1190C, held for 12 hours, then cooled for 100 hours to 800C (80K below the melting point of TmAg$_2$ measured from a differential scanning calorimeter), held there for another 100 hours to increase crystal quality, and then cooled to room temperature. The resulting melt had large sections with crystalline facets/features which could be broken from the boule and oriented with single crystal X-ray diffaction.

Heat capacity measurements were performed in a Quantum Design PPMS using a standard thermal relaxation technique.  Apiezon N-grease was used to secure the sample to the sapphire platform. A correlation was also noted between the measured resistance ratio at 300K vs 2K (RRR) and the sharpness of the phase transition signature in heat capacity, as well as the ratio of the magnitude of the critical feature to the non-critical background. Only the measurements of samples with the highest ratio of critical heat capacity to background and RRR greater than 50 were measured for heat capacity and elastocaloric measurements.

For the elastocaloric measurements, a commerical Razorbill CS-100 strain cell was used to apply strain to the samples, which were polished and cut along the (1 0 0) direction via wire-saw to be approximately 2mm x 0.04mm x 0.06mm in size. The samples were secured between two sets of mounting plates using Stycast 2850FT Epoxy, which were screwed into the strain cell, to have a gap of approximately 1mm. An AC voltage of 20V RMS at 72 Hz was applied to the outer piezoelectric (PZT) stacks of the strain cell, which corresponded to applying an AC displacement of the sample of approximately 0.01\% of its length. This frequency was experimentally determined by measuring the elastocaloric signal at 10K for frequencies in the range of 10-200 Hz, and choosing the frequency with the largest response. This implied the frequency was at the plateau of the relevant thermal transfer function, which did not observably shift in the temperature range measured~\cite{ikeda2019ac}. DC voltages were applied to both the outer set of PZTs and the inner PZT to reach a strain range of 0.3\% at low temperatures. To approximate the strain the sample experienced a capacitor built in the strain cell was measured to provide the relative displacement of the sample plates, which was divided by the length of the gap. This however only approximated $\varepsilon_{xx}$ of the sample as it assumes a 100\% strain transmission, and so we denote this quantity as $\varepsilon'_{xx}$ for the remainder of the paper.

The temperature fluctuations in the sample induced by the AC strain were measured using a small commercial RuO$_2$ thin-film resistor/thermometer (Vishay \# CRCW01002K43FREL) which was attached to the sample via a 50$\mu$m diameter gold wire. The gold wire was silver pasted to the sample as well as the back of the thermometer, to provide high thermal conductivity from the thermometer to the sample. The ceramic substrate of the thermometer was polished (not affecting the thin-film) to have dimensions of 150$\mu$m x 75$\mu$m x 20$\mu$m to reduce its heat capacity. The temperature signal was measured at the sum of this strain frequency and the thermometer excitation frequency, extracted with an SRS860 lock-in using a similar AC demodulation technique as described in Ref.~\cite{Alex}. To increase the signal to noise ratio the voltage from the thermometer was measured using a custom Wheatstone bridge in which the 3 other resistive components were identical resistors to the thermometer, and the bridge was thermally anchored to the cell.

\section{Acknowledgements}
We want to acknowledge fruitful conversations with E. Berg. This work was supported by the Gordon and Betty Moore Foundation Emergent Phenomena in Quantum Systems Initiative through Grant GBMF9068. 

\section{Author Contributions}
 E.R., M.I, conducted the measurements. E.R performed the calculations. E.R grew the samples. I.R.F. oversaw the project. E.R., and I.R.F. wrote the manuscript with input from all authors.

\section{Competing Interests}
The authors declare no competing interests.

\section{Data Availability}
All data supporting the findings of this study are available upon request.

\newpage

\bibliography{main}

\end{document}